\documentclass[aps,prl,reprint,groupedadresses]{revtex4-1}

\usepackage{graphicx}
\usepackage[utf8]{inputenc}
\usepackage{amsmath, amssymb}
\usepackage{hyperref}

\newcommand{\fig}[1]{Fig.\,\ref{#1}}
\newcommand\rvec{\mathbf{r}}
\newcommand\kvec{\mathbf{k}}
\newcommand\xvec{\mathbf{x}}
\newcommand\yvec{\mathbf{y}}
\newcommand\lth{\lambda_\mathrm{th}}

\date{\today}
\begin{document}

\title{Momentum distribution in the unitary Bose gas from first principles}

\author{Tommaso Comparin}
\email{tommaso.comparin@ens.fr}
\author{Werner Krauth}
\email{werner.krauth@ens.fr}
\affiliation{Laboratoire de Physique Statistique, \'{E}cole Normale
Sup\'{e}rieure/PSL Research University, UPMC, Universit\'{e} Paris
Diderot, CNRS, 24 rue Lhomond, 75005 Paris, France}


\begin{abstract}
We consider a realistic bosonic $N$-particle model with unitary
interactions relevant for Efimov physics.  Using quantum Monte Carlo
methods, we find that the critical temperature for Bose-Einstein
condensation is decreased with respect to the ideal Bose gas.  We also
determine the full momentum distribution of the gas, including its
universal asymptotic behavior, and compare this crucial observable to
recent experimental data.  Similar to the experiments with different
atomic species, differentiated solely by a three-body length scale, our
model only depends on a single parameter.  We establish a weak influence
of this parameter on physical observables.  In current experiments,
the thermodynamic instability of our model from the atomic gas towards
an Efimov liquid could be masked by the dynamical instability due to
three-body losses.
\end{abstract}

\maketitle 

First predicted in 1970\cite{Efimov1970PLB}, the Efimov effect describes
the behavior of three strongly interacting bosons when any two of them
cannot bind. At unitarity, when the scattering length diverges, the
three-body bound states are scale invariant and they form a sequence up
to vanishing binding energy and infinite spatial extension.  Efimov
trimers had been intensely discussed in nuclear physics, but it was
in an ultracold gas of caesium atoms that they were finally
discovered\cite{Kraemer2006}.
To observe Efimov trimers, experiments in atomic physics rely on
Feshbach resonances\cite{Chin2010RMP}, which permits to instantly
switch a gas between weak interactions and the unitary limit.
Such a control of interactions lacks in nuclear physics or condensed
matter experiments, and singular interactions can be probed there only
in the presence of accidental fine tuning\cite{Braaten2006PR}.
Beyond the original system\cite{Kraemer2006}, Efimov trimers have
now been observed for several multi-component systems,
including bosonic, fermionic and Bose-Fermi mixtures
\cite{Barontini2009PRL, Williams2009PRL, Pires2014PRL}.
These experimental
findings are interpreted in terms of the theory of few-body strongly
interacting quantum systems. For three identical bosons in three
dimensions, a complete universal theory is available, on and off
unitarity \cite{Braaten2006PR}. Further theoretical work is aimed
at understanding bound states for more than three bosons, mixtures,
and the effects of dimensionality.

\begin{figure}[t]
\centering
\includegraphics[width=\linewidth]{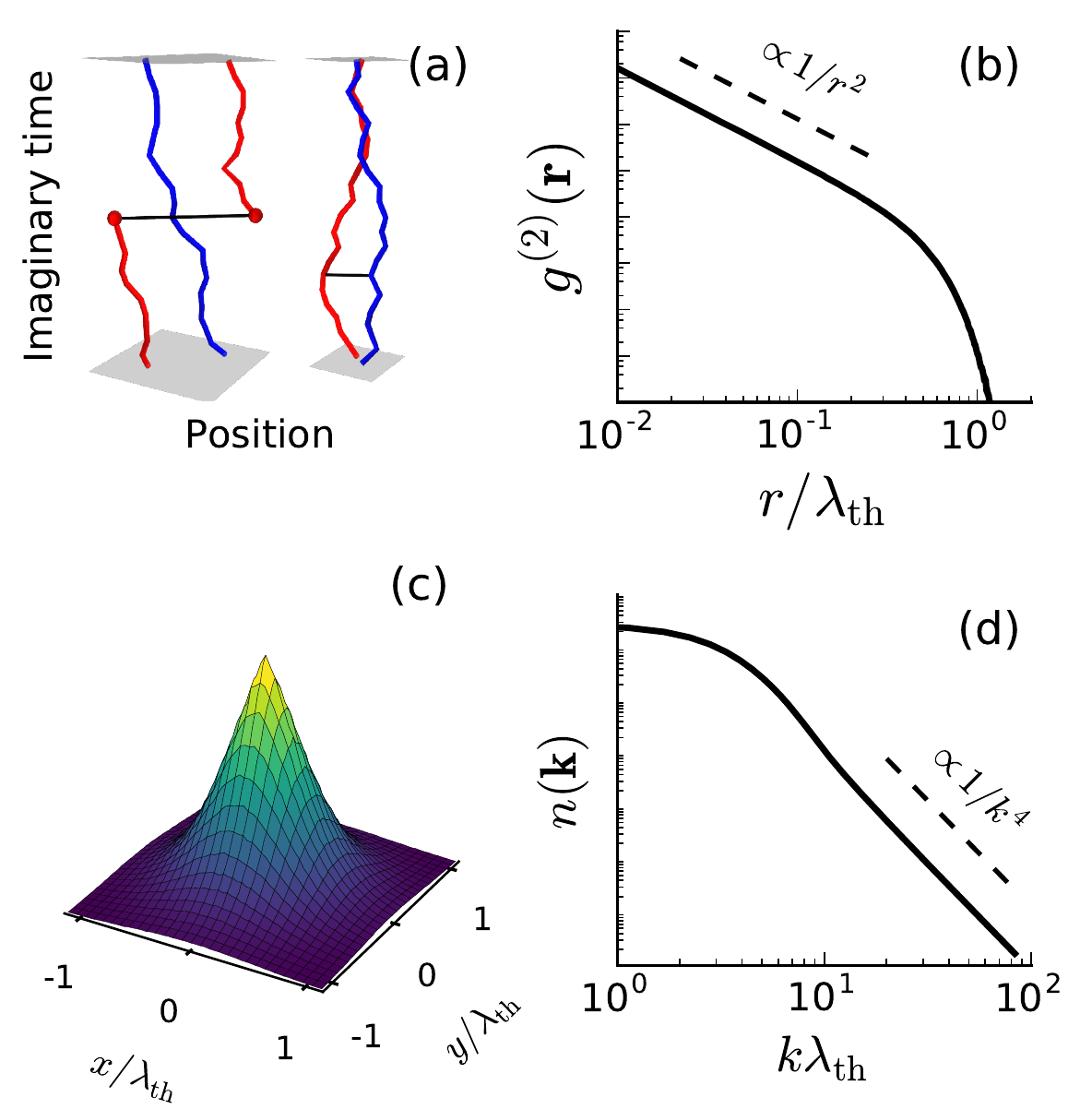}
\caption{
Correlation functions for two unitary bosons.
(a)
Open (\emph{left}) and  closed (\emph{right}) co-cyclic configurations
in the path-integral representation. Closed
configurations yield $g^{(2)}(\rvec)$. Open configurations yield
$n(\kvec)$ and its inverse Fourier transform $g^{(1)}(\rvec)$.
(b)
Pair-correlation function $g^{(2)}(\rvec)$ (distance distribution in closed
configurations), featuring a $r^{-2}$ divergence at small
$r$.
(c)
Cut of $g^{(1)}(\rvec)$ (distribution of the distance between open ends),
for $\rvec = (x, y, 0)$,
illustrating the  cusp at $\rvec \simeq 0$.
(d)
Momentum distribution $n(\kvec)$ with
asymptotic decay, $ \propto 1 / k^4$, at large $\kvec$.
\label{fig:1}}
\end{figure}

Near-unitary interparticle interactions also impact the thermodynamics
of the atomic gas, the description of which presents a challenge
beyond the traditional theory of the Efimov effect. In addition, mean-field
theory does not apply to infinite interactions\cite{Lee1957PR}, and the virial
expansion\cite{Castin2013CJP} fails to describe the low-temperature state.
Moreover, in atomic-physics experiments, strong interactions enhance the
three-body loss rate, making the gas of bosons unstable.
A characterization of the universal dynamics of these losses has been
recently achieved\cite{Rem2013PRL, Fletcher2013PRL, Eismann2015PRX}.
On the other hand, a single breakthrough experiment\cite{Makotyn2014NP}
has addressed the low-temperature thermodynamics for a unitary bosonic
gas, coming to the conclusion that equilibrium was approached faster
than the system life-time. The importance of this system stems from its
universal character: All differences between atomic species may be encoded
in a single three-body parameter, related to the van der Waals length
\cite{Wang2012PRL}. However, this prominent experiment could not be interpreted
univocally, as first-principle theoretical predictions were lacking.
In the present work, we obtain these predictions for a model which shares the
experimental system's universality.
We develop a novel quantum Monte Carlo algorithm\cite{Comparin2016b}, that
overcomes the peculiar challenges posed by the infinite interactions. This
allows us to compute
the critical temperature for Bose-Einstein condensation, and the full momentum
distribution $n(\kvec)$ throughout the entire phase diagram, including its
universal asymptotic behavior.

In the unitary limit, the scattering length diverges, and atomic pair
interactions are powerful yet very short-ranged.  The bosonic pair
correlation function $g^{(2)}(\rvec)$ diverges as $1/r^2$ at short
distances $r = | \rvec |$, yet two isolated unitary bosons barely
hold together: They form a molecule of infinite radius and vanishing
binding energy.  In thermodynamic equilibrium, three or more such
bosons, with zero-range interactions, collapse into a single point,
unless the unitary pair interactions are counterbalanced by a three-body
repulsion. In experimental systems the latter is effectively realized by
the van der Waals potential\cite{Wang2012PRL}, so that the unitary Bose
gas is stabilized against collapse.  The divergence of $g^{(2)}(\rvec)$
persists in the gas, with a finite contact density $c_2 = \lim_{r \to 0}
(4 \pi r)^2 g^{(2)}(\rvec)$.  The large-$k$ asymptotics of the momentum
distribution\cite{Tan2008APa, Werner2012PRA_2} is governed by Tan's
contact parameter $C_2 = c_2 V$ (where $V$ is the system volume), and
it decays as $n(\kvec)\simeq C_2 / k^4$ for $k \to \infty$.

We consider $N$ bosons at temperature $T$ in a periodic cubic
box (thermodynamic $NVT$ ensemble). Pair interactions are of zero range
and infinite depth, and the resonant two-body bound state realizes an
infinite scattering length. In addition, any three particles $a, b, c$
are subject to a hard cutoff $R>R_0$ on their hyperradius $R$, defined
as the mean of their squared pair distances: $R^2 \equiv(r^2_{ab} +
r^2_{bc} + r^2_{ac}) / 3$.  This realistic model describes ultracold
atomic ensembles with an interaction range much smaller than the
scattering length, the interparticle distance and the thermal de Broglie
wavelength.  The two-body interactions, with their infinite scattering
length, provide no scale. The model's phase diagram thus depends
on two dimensionless numbers, namely the thermal de Broglie wavelength
$\lth \rho^{1/3}$, and the three-body cutoff $R_0 \rho^{1/3}$, both in
units of the typical interparticle distance $\rho^{-1/3}$ (where $\lth
= \sqrt{2\pi \hbar^2 \beta /m}$, $\beta=1/(k_B T)$, and $\rho = N/V$).
At high temperature, three-particle effects are suppressed, and the model
depends only on $\lth \rho^{1/3}$.  In experiments at low temperature,
three-body correlations lead to strong recombination losses, with a
loss rate scaling as $\sim T^{-2}$\cite{Rem2013PRL, Fletcher2013PRL,
Eismann2015PRX}, the predominant source of instability of the system. In
contrast, our model conserves particle number.

Path-integral quantum Monte Carlo techniques allow us to solve this
model from first principles, that is, without systematic errors.
Computational challenges are posed by the divergence of $g^{(2)}(\rvec)$
at contact (see \fig{fig:1}(b)) and by the need to determine $n(\kvec)$
for large momenta $k$ (see \fig{fig:1}(d)).  This corresponds to
computing the single-particle correlation function $g^{(1)}(\rvec)$
-- the inverse Fourier transform of $n(\kvec)$ -- at small $r$,
close to its cusp singularity at $r\to0$ (see \fig{fig:1}(c)).
Our path-integral quantum Monte Carlo algorithm\cite{Ceperley1995RMP,
Krauth1996PRL, Boninsegni2006PRE, Comparin2016b} samples both closed
and open path-integral configurations (\emph{cf.} \fig{fig:1}(a)).
The former give access to the superfluid fraction $\rho_s/\rho$ (via the
winding-number estimator\cite{Pollock1987PRB}) and to the pair-correlation
function $g^{(2)}(\rvec)$ (from which we extract the contact density $c_2$).
 Open configurations, in contrast,
sample the single-particle correlation function $g^{(1)}(\rvec)$, and
give access to the normalized momentum distribution (satisfying
$\int d\kvec\,n(\kvec)/(2\pi)^3=N$ in the normal gas).
A dedicated estimator allows us to sample $n(\kvec)$ for arbitrarily
large momenta $k$ (\emph{cf.} Supp. Item 1 \cite{SuppComparin2016a}).

\begin{figure}[tb]
\centering
\includegraphics[width=\linewidth]{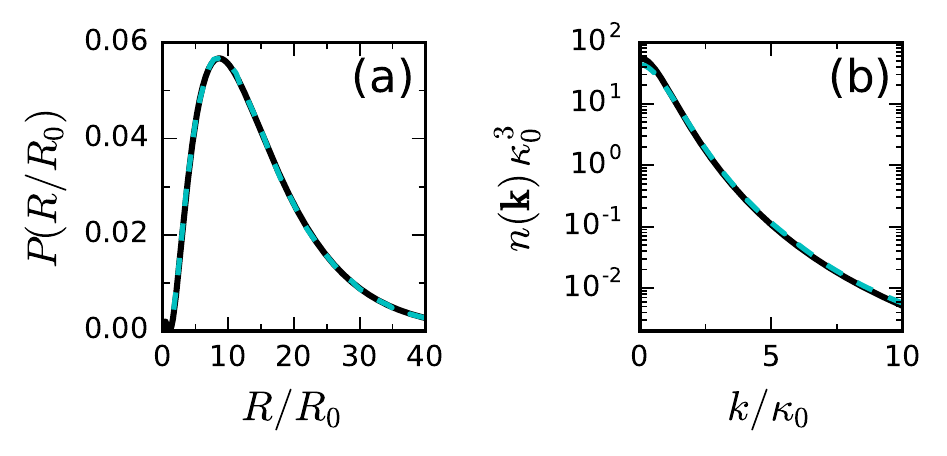}
\caption{
Correlation functions for three unitary bosons.
(a)
Hyperradial probability distribution for three co-cyclical bosons with 
hyperradial cutoff at low temperature (\emph{cyan dashed line})
and for the universal Efimov
trimer (\emph{black solid line}, from Ref.\,\cite{Braaten2006PR}).
(b)
Momentum distribution for three co-cyclical bosons
(\emph{cyan dashed line}), and for the universal trimer
(\emph{black solid line}, from Ref.\,\cite{Castin2011PRA}), in units of the
trimer binding momentum $\kappa_0$.
\label{fig:2}}
\end{figure}

We include zero-range unitary interactions between two bosons through the
exact two-body propagator\cite{Piatecki2014NC, Yan2015PRA_1}, and treat
them with a highly efficient direct-sampling approach\cite{Comparin2016b}.
The many-body density matrix is then built via the pair-product
approximation.  The hyperradial cutoff is included via the
Trotter break-up\cite{Ceperley1995RMP}, and an effective value of $R_0$
is obtained -- for a finite imaginary-time discretization -- through the
comparison with the expression for the hyperradial wave function of a
single universal trimer\cite{Piatecki2014NC, Comparin2016b}.
For three unitary bosons, the length scale $R_0$ sets a lower bound on
the Efimov energy spectrum, and specifies a three-body ground state. At
low temperature, our Monte Carlo simulations for $N=3$ allow us to obtain
excellent agreement of the hyperradial probability distribution and the
momentum distribution for our model with the corresponding 
quantities for the universal Efimov trimer \cite{Braaten2006PR,
Castin2011PRA} (see \fig{fig:2}(a) and \fig{fig:2}(b)), providing also
a parameter-free check of our computer program.

\begin{figure*}[bt]
\centering
\includegraphics[width=\linewidth]{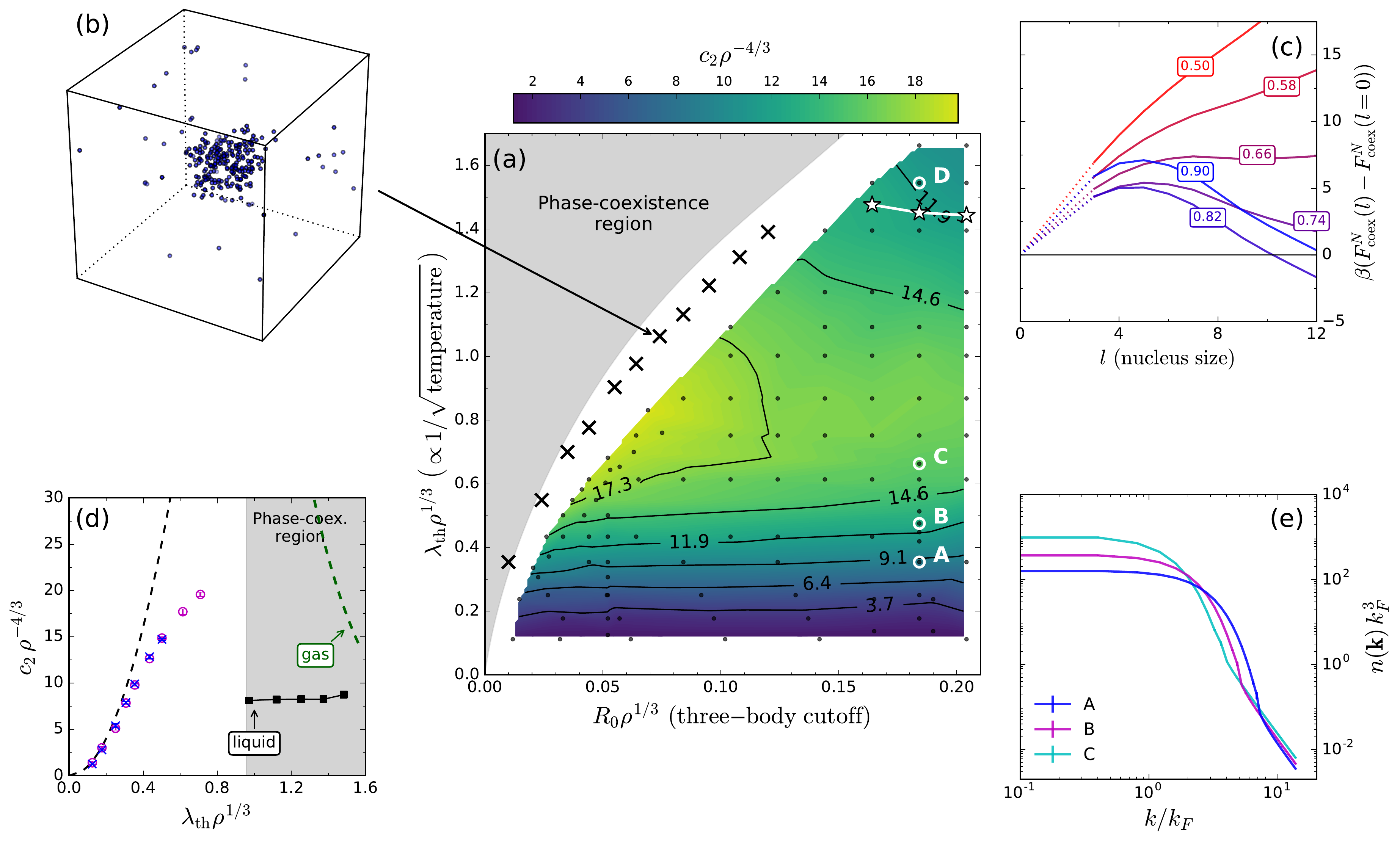}
\caption{
Equilibrium phase diagram of unitary bosons.
(a)
Contact density $c_2 \rho^{-4/3}$, as a linear interpolation of
numerical results (extracted from 
$g^{(2)}(\rvec)$, for $N=64$).
\emph{White stars}: transition between normal gas and superfluid 
(Bose-condensed) phase.
\emph{Black crosses}: Phase-separated points.
\emph{Gray area}: Phase-coexistence region \cite{Piatecki2014NC}.
(b)
Stable Efimov-liquid droplet coexisting with a normal gas ($N = 
256$).
(c)
Excitation free energy for the Efimov-liquid nucleation, \emph{vs.}
nucleus size $l$.
$\lth\rho^{1/3}$ varies between lines (see labels), between $0.5$ (monotonically increasing,
\emph{red line}) and $0.9$ (barrier, \emph{blue line}). The hyperradial cutoff
is fixed ($R_0 \rho^{1/3}=0.03$), and the phase-separation region sets in at
$\lth\rho^{1/3}\simeq0.66$.
(d)
Contact density $c_2\rho^{-4/3}$ \emph{vs.} $\lth\rho^{1/3}$, for $R_0
\rho^{1/3}=0.052$:
Virial expansion (\emph{black dashed line}) and numerical results, via the
$n(\kvec)$ and $g^{(2)}(\rvec)$ estimators (\emph{crosses, circles}).
In the phase-coexistence region,
the liquid and gas phases have different contact densities
(for the gas, the virial expansion is used).
(e)
Momentum distribution (in units of the Fermi momentum $k_F =
(6\pi^2\rho)^{1/3}$) for parameters corresponding to points A, B, and C, in
panel (a).
\label{fig:3}}
\end{figure*}

In the thermodynamic $NVT$ ensemble, unitary bosons phase-separate below a 
given
temperature into a normal or Bose-condensed gas dominated by entropy and a 
high-density Efimov liquid of low potential energy (see \fig{fig:3}(a) and
Ref.\,\cite{Piatecki2014NC}).
An equilibrium liquid bubble 
forms inside the gas (\fig{fig:3}(b)), and
the nucleation process is reversible across the coexistence line. For $R_0 \to
0$, the gas-to-liquid condensation energy $\propto 1/ R_0^2$
overcomes the gas entropy loss at arbitrarily high 
temperatures, so that the coexistence line starts at $\lth \rho^{1/3} = R_0
\rho^{1/3} = 0$. 
In the phase-coexistence region, the free energy $F^N_\mathrm{coex}(l)$ is
composed of two terms, corresponding to the Efimov-liquid nucleus of $l$
particles and to the gas of the remaining $N-l$ particles. An analytical model,
based on the virial expansion of the gas\cite{Castin2013CJP} and the known
ground-state energies for small Efimov-liquid nuclei\cite{VonStecher2010JPB}
(supposed incompressible), allows us to model the excitation free energy (see
Supp. Item 5 \cite{SuppComparin2016a}).
In the homogeneous gas phase, $F^N_\mathrm{coex}(l)$
monotonically increases with $l$ (\fig{fig:3}(c)). At lower temperatures, the
gas becomes metastable, with a free-energy barrier at a critical cluster size
$l^*$. The nucleation rate per volume is proportional to $\exp(-\beta \Delta
F)$, where $\Delta F = F^N_\mathrm{coex}(l^*)- F^N_\mathrm{coex}(0)$ is the
free-energy barrier to overcome the critical cluster size $l^*$.
At low temperature, $\beta \Delta F$ and therefore the
characteristic nucleation time for the Efimov liquid remain
finite (see \fig{fig:3}(c)).
The observed long experimental life-time \cite{Makotyn2014NP} is consistent with
the idea that the phase-separation instability does not take place, in current
experiments, as
three-body losses effectively destabilize liquid droplets before
the critical nucleus size $l^* \simeq 5$ is reached.
A study of the many-body quantum dynamics will be needed to confirm this
hypothesis.

In the stable region of the phase diagram, the momentum distribution 
$n(\kvec)$ is in principle obtained as the Fourier transform 
of $g^{(1)}(\rvec)$, the distribution for distance vectors of open 
configurations (\emph{cf.} \fig{fig:1}(a)). This estimator, however, poorly 
samples the
short-distance cusp in $g^{(1)}(\rvec)$ (equivalently, the large-$k$ behavior 
of $n(\kvec)$).
Our approach is rather based on an average of the analytical two-body
expression, to determine $n(\kvec)$ at arbitrarily large $k$ (see 
Supp. Item 1 \cite{SuppComparin2016a}).
The asymptotic behavior of $n(\kvec) = C_2/k^4$ for $k \to 
\infty$ is also contained in the contact density, obtained from closed-path 
configurations (see \fig{fig:3}(d)).
In the normal phase, the small-$k$ part of the momentum distribution $n(\kvec)$
resembles the one of ideal bosons:
The peak at $k=0$ corresponds to the Maxwell-Boltzmann distribution $\exp(-\beta
k^2/2)$ in the classical limit (at high temperature), and the narrowing at lower
temperature is enhanced by bosonic statistics (see \fig{fig:3}(e)).
At large $k$, $n(\kvec)$ crosses over into the $C_2/k^4$ asymptotic behavior,
with a crossover point which scales as $k/k_F\propto1/(\lth\rho^{1/3})$ for
large temperature,
where $k_F \equiv (6\pi^2\rho)^{1/3}$ is the Fermi momentum.
In the phase-coexistence region, we
obtain two different contact  densities for the gas and for the Efimov
liquid (see \fig{fig:3}(d)).

Throughout the homogeneous region, the momentum distribution only depends weakly
on $R_0 \rho^{1/3}$, both in the full $n(\kvec)$ and in its asymptotic tail,
underlining the generality of the model under study.
The slow decrease of $c_2$ for increasing $R_0 \rho^{1/3}$ (absent at high
temperature,
$\lth \rho^{1/3} \to 0$) corresponds to a small suppression of $g^{(2)}(\rvec)$
at short distance, indirectly caused by the hyperradial cutoff.
At high temperature, our first-principles results for the contact density  
rapidly fall  below the
predictions of the virial expansion\cite{Smith2014PRL, Barth2015PRA,
Liu2015PRA} (\fig{fig:3}(d)), leveling off at intermediate temperature, and 
finally
decreasing at lower temperature. This non-monotonic behavior was already
qualitatively predicted\cite{Liu2015PRA}. The low-temperature values of
$c_2\rho^{-4/3}$ fall in the same range of previous zero-temperature approximate
results\cite{Diederix2011PRA, Rossi2014PRA, Sykes2014PRA} (\emph{cf.} Supp. Item
2 \cite{SuppComparin2016a}).

\begin{figure}[t]
\centering
\includegraphics[width=\linewidth]{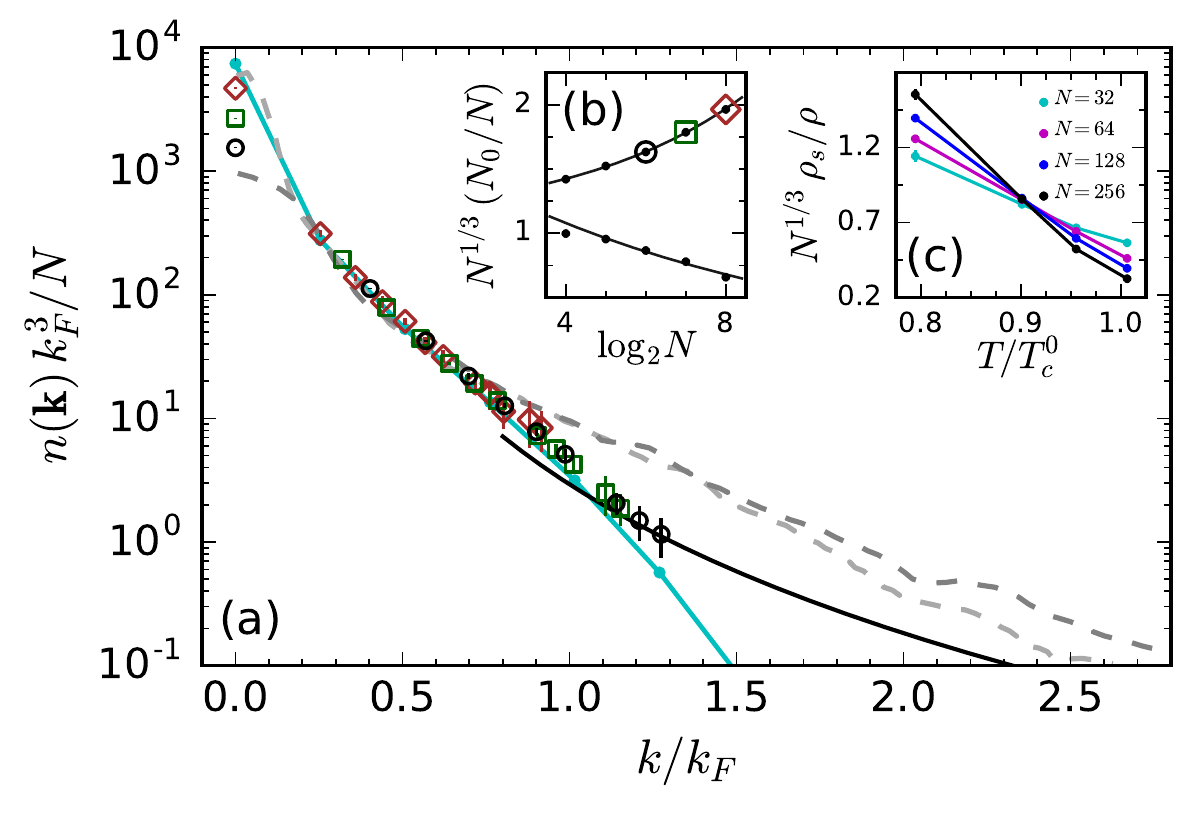}
\caption{
Full momentum distribution $n(\kvec)$ in the Bose-condensed gas phase.
(a)
$n(\kvec)$ at $\lth \rho^{1/3}=1.545$, $R_0\rho^{1/3} = 0.184$ (point D in
\fig{fig:3}(a)). First-principles results for $N=64, 128,$ and $256$ (\emph{black
circles}, \emph{green squares}, \emph{brown diamonds}, respectively), and
$\propto C_2/k^4$ asymptotic behavior for $k \to \infty$ (for $N=64$,
\emph{black solid line}). Dashed lines are experimental data of
Ref.\,\cite{Makotyn2014NP} for two different densities. The momentum
distribution for $N=256$ ideal bosons is also shown (\emph{cyan solid line}).
(b)
Scaling of the condensed fraction $N_0/N$ with the system size, in the normal
and condensed phases. The upper curve (at $T < T_c$) corresponds to the 
parameters in panel (a), and the exact numerical data are fitted by $N^{1/3}
(N_0/N) \simeq 1.06 + 0.14 N^{1/3}$ (same symbols for $N$ as in panel (a)).
The lower curve is at $\lth \rho^{1/3}\simeq1.373$ (corresponding to $T > T_c$),
and is fitted by $ N^{1/3} (N_0/N) \simeq1.71 N^{ -1/6}$.
(c)
Rescaled superfluid fraction \emph{vs.} temperature, at $R_0\rho^{1/3}=0.184$.
The crossing point at $T/T_c^0\simeq0.9$ (corresponding to
$\lth\rho^{1/3}\simeq1.45$) shows that $T_c$ is lowered by $10\% $ with respect
to the ideal Bose gas, in the limit $N \to \infty$.
}
\label{fig:4}
\end{figure}

For large three-body cutoff ($R_0\rho^{1/3} \gtrsim 0.16$), the normal gas 
Bose-condenses before phase separation sets in.
At finite $k$, $n(\kvec)$ has very small
finite-size effects, making the comparison with experiments feasible.
Data for harmonically trapped Rb atoms\cite{Makotyn2014NP} are available up
to $k/k_F \simeq 3$
and they are considered equilibrated for $k/k_F \gtrsim 0.5$.
At small $k$, the harmonic-trap geometry has the strongest influence.
Up to momenta $k \approx k_F$, the experimental curves overlap with the
theoretical data (see \fig{fig:4}(a)).
As the asymptotic $k^{-4}$ behavior of $n(\kvec)$ sets in for the numerical
curve ($k \gtrsim 1.1k_F$, at the chosen temperature), the experimental curve
remains higher.
This deviation is difficult to reconcile with our model, as
the $k^{-4}$ prefactor is expected to decrease even further at lower
temperature (see Supp. Item 2 \cite{SuppComparin2016a}).

The condensate fraction is related to the $\kvec = \mathbf{0} $ component of $n(\kvec)$,
$N_0/N = n(\kvec=\mathbf{0})/(N V)$.  Below the critical
temperature $T_c$, it remains non-zero for $N \to \infty$, with finite-size
corrections $\propto N^{- 1/3}$.
In the normal phase, in contrast, the large-$N$ limit of $N_0/N$ vanishes.
These two behaviors are clearly distinguished in the data (see \fig{fig:4}(b)).
The scaling of the superfluid fraction
yields a precise estimate of the critical temperature\cite{Pollock1992PRB} (see
\fig{fig:4}(c) and Supp. Item 3 \cite{SuppComparin2016a}).
In the unitary Bose gas, $T_c$ is reduced  by $ 10 \%$: The critical
value of $\lth \rho^{1/3}$ is between $1.44$ and $1.48$ (see \fig{fig:3}(a)),
while the ideal-bosons transition\cite{Pitaevskii2003} is at
$\lth\rho^{1/3}\simeq1.377$.
The deviation of $T_c$ from $T_c^0$ (the ideal-bosons critical temperature) is
larger for smaller values of $R_0 \rho^{1/3}$.
It is instructive to compare $n(\kvec)$ with the ideal-gas curve.
Unitary interactions cause a depletion of the condensate, i.e. a decrease of
$N_0/N$. At temperature $T \lesssim T_c$, this follows from the
negative shift of the critical temperature, $T_c < T^0_c$.
While the $\kvec=\mathbf{0}$ component of $n(\kvec)$ is smaller, on the other hand, the
tail
of the distribution is more important (\emph{cf.} the power-law $k^{-4}$ decay,
\emph{vs.} the exponential suppression of $n(\kvec)$ for ideal
bosons).
At point D in \fig{fig:3}(a), the depletion of the condensate is not entirely
compensated by the large-$k$ contribution
(see Supp. Item 4
\cite{SuppComparin2016a}).
This leads to the reweighting of the unitary gas momentum distribution
with respect to the one of the ideal Bose gas, without introducing any new 
features. 

Both for three-body and many-body states of unitary bosons, $n(\kvec)$ has
subleading oscillations around the $C_2/k^4$ asymptotic tail. These consist in a
log-periodic function of $k$, modulated by $C_3/k^5$\cite{Castin2011PRA,
Braaten2011PRL}.
The three-body contact parameter $C_3$
vanishes at the length scale of the short-range hyperradial repulsion between
atoms, induced by the van der Waals potential\cite{Wang2012PRL} or by the
explicit hyperradial cutoff $R_0$.  Thus the subleading oscillations can in our
model only be observed for $k \lesssim 1/R_0$.
For our high-temperature results (\emph{cf.} \fig{fig:3}(e)), the asymptotic
tail of $n(\kvec)$ kicks in at $k > 1/R_0$, where $C_3$ is effectively zero, and
we do not expect visible subleading corrections. At low temperature,
however, the crossover into the asymptotic tail is at $k \approx 1/R_0$ (see
\fig{fig:4}(a)). Thus the subleading oscillations are possibly observable within
the three-body-cutoff model, for a slightly smaller value of $T$ or $R_0$,
despite being beyond the current resolution.

In conclusion, we have computed the equilibrium phase diagram and the momentum 
distribution of the unitary Bose gas from first principles, overcoming the
technical challenges through a novel quantum Monte Carlo
algorithm\cite{Comparin2016b}.
Our theoretical predictions will most easily be checked in the currently
available homogeneous
traps\cite{Gaunt2013PRL, Navon2015}, which
are less subject to three-body losses than the traditional harmonic traps.
In the near future, we expect high-precision experimental measurements of the 
superfluid transition and of the momentum distribution $n(\kvec)$ in the unitary 
Bose gas.

\begin{acknowledgments}
We thank Riccardo Rossi for insightful suggestions, and acknowledge extensive
discussions with Kris van Houcke, Xavier Leyronas and F\'{e}lix Werner. We thank
Yvan Castin and Eric Cornell for discussions, and for allowing us reuse of data
in Ref.\,\cite{Castin2011PRA} and in Ref.\,\cite{Makotyn2014NP}.  This work was
performed in part at the Aspen Center for Physics, which is supported by
National Science Foundation grant PHY-1066293.  This work was granted access to
the HPC resources of MesoPSL financed by the Region Ile de France and the
project Equip@Meso (reference ANR-10-EQPX-29-01) of the programme
Investissements d'Avenir supervised by the Agence Nationale pour la Recherche.
\end{acknowledgments}

\bibliographystyle{apsrev4-1}
\bibliography{comparin_krauth_2016a}

\clearpage
\onecolumngrid
\renewcommand{\theequation}{S\arabic{equation}}
\renewcommand{\thefigure}{S\arabic{figure}}
\setcounter{equation}{0}
\setcounter{figure}{0}
\setcounter{page}{1}

\subsection{Supplementary Item 1: Momentum-distribution estimator}

\begin{figure}[h!]
\centering
\includegraphics[width=0.4\linewidth]{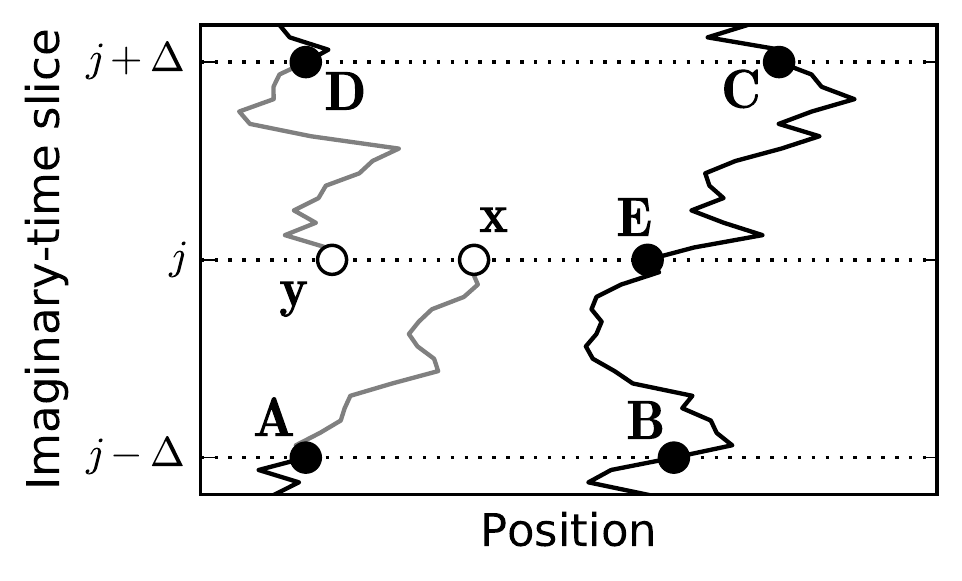}
\caption{
Open $N=2$ path configuration.
An estimator for the momentum distribution is derived from 
the analytical expression for $n(\kvec)$ for fixed
positions
$ \mathbf{A}, \mathbf{B}, \mathbf{C}, \mathbf{D}, \mathbf{E}$.
\label{suppfig:ABCDE}}
\end{figure}

In path-integral quantum Monte Carlo, the momentum distribution is usually
computed from the exponential $ e^{ -i \kvec \cdot (\xvec - \yvec)} $ (with open
ends $\xvec$ and $\yvec$), averaged over open-path
configurations\cite{Ceperley1995RMP, Boninsegni2006PRE}. At large momenta $k$ --
where $n(\kvec)$ tends to zero -- this estimator becomes unpractical,
because of a vanishing signal-to-noise ratio.
We construct a new estimator (used in Fig.\,3(e)), based on the solution
of the two-body problem represented in
\fig{suppfig:ABCDE}\cite{Comparin2016b}.
We analytically determine 
$n(\mathbf{k}| \mathbf{A}, \dots, \mathbf{E})$, the average of
$e^{-i\kvec\cdot(\xvec-\yvec)}$ for given positions 
$ \mathbf{A}, \mathbf{B}, \mathbf{C}, \mathbf{D}, \mathbf{E}$.
For $N=2$, $n(\kvec)$ is obtained as an average 
of 
$n(\kvec| \mathbf{A}, \dots, \mathbf{E})$
over configurations $ \mathbf{A}, \mathbf{B}, \mathbf{C}, \mathbf{D}, 
\mathbf{E}$ sampled during the simulation. 
For $N\ge3$, this coarse-grained estimator holds for ``local''
configurations, where the two open ends are close to each other  
and to the nearest of the other bosons ($\xvec \sim \yvec \sim 
\mathbf{E}$). For non-local configurations 
we again resort to the direct estimator
$\left \langle
e^{-i \kvec \cdot (\xvec - \yvec)}
\right\rangle$,
and finally obtain $n(\kvec)$ as a weighted average of the two 
estimators (\emph{cf.} Ref.\,\cite{Comparin2016b}). This procedure relies on an 
appropriate cutoff between local and non-local configurations. 
At high enough temperature, where the procedure is used, we carefully check that
the contact density $c_2$ extracted from the asymptotic behavior of $n(\kvec)$
for $k \to \infty$ agrees with the $r \to 0$ limit of $g^{(2)}(\rvec)$ (see
Fig.\,3(c)).

\subsection{Supplementary Item 2: Contact density at low temperature}
\begin{figure}[hbt]
\centering
\includegraphics[width=0.5\linewidth]{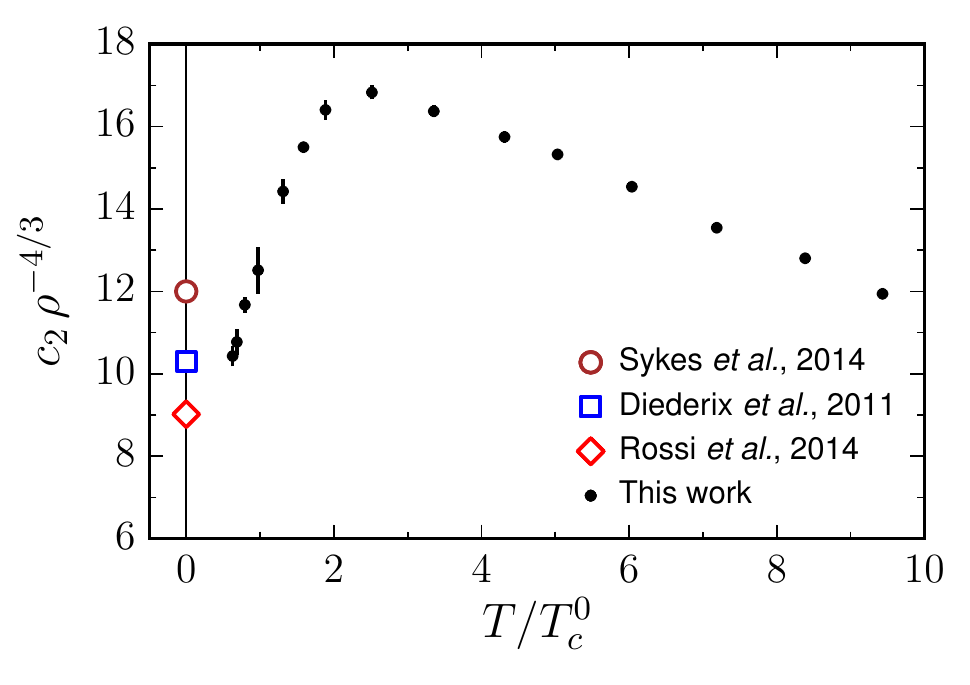}
\caption{
Contact density at low temperature ($T/T_c^0=1$ corresponds to $\lth
\rho^{1/3}=1.377$), for $R_0 \rho^{1/3} = 0.184$ (\emph{black points}), and
zero-temperature approximate results for the models in
Refs.\,\cite{Diederix2011PRA, Rossi2014PRA, Sykes2014PRA} (\emph{open
symbols}).
\label{suppfig:c2_at_low_T}}
\end{figure}

In \fig{suppfig:c2_at_low_T}, the  
data of Fig.\,3(a) are plotted as a function of $T / T_c^0$ , for a 
three-body cutoff $R_0\rho^{1/3} \simeq 0.184$.
Our first-principles low-temperature values for $c_2 \rho^{-4/3}$ are roughly
compatible with the zero-temperature approximate results from
Refs.\,\cite{Diederix2011PRA, Rossi2014PRA, Sykes2014PRA}. These are obtained
via a Jastrow ansatz and hypernetted-chain approximation\cite{Diederix2011PRA},
a quantum Monte Carlo calculations based on a Jastrow-Feenberg
ansatz\cite{Rossi2014PRA}, and a time-dependent variational ansatz for the
many-body state\cite{Sykes2014PRA}.
The value $c_2 \rho^{-4/3}\simeq 22$, extracted from an analysis of the
experimental data\cite{Smith2014PRL}, appears significantly larger than
our theoretical predictions.

\subsection{Supplementary Item 3: Superfluid transition}
\begin{figure}[hbt]
\centering
\includegraphics[width=0.97\linewidth]{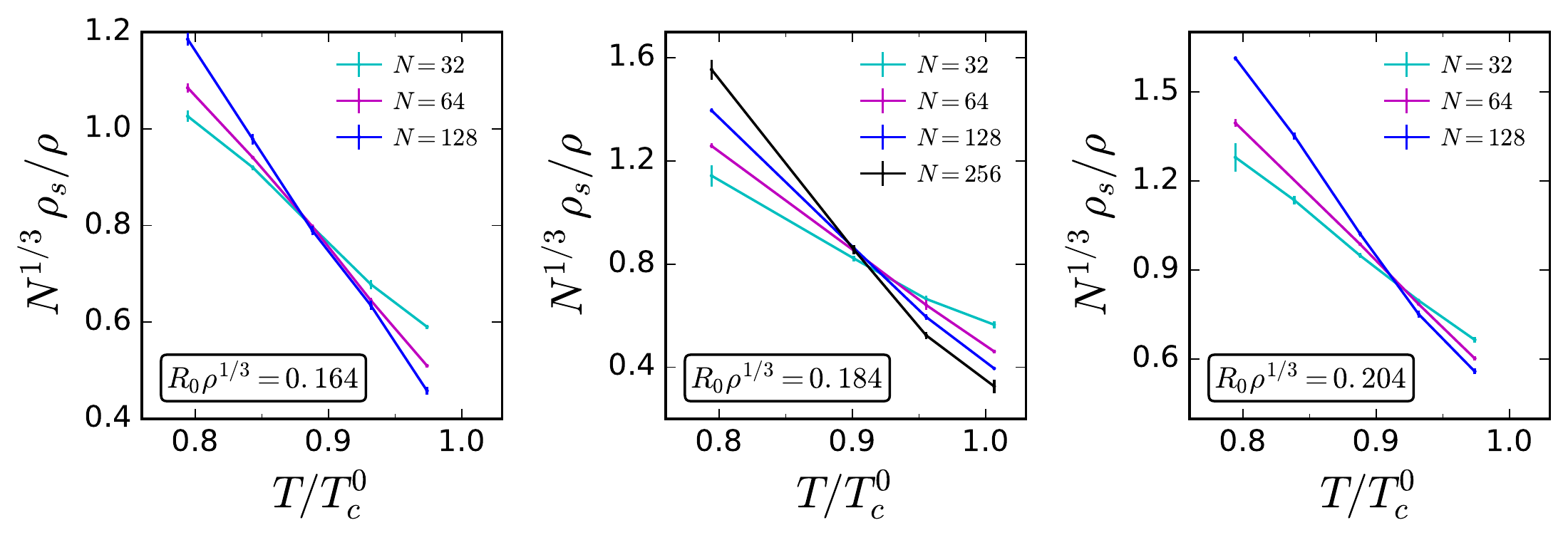}
\caption{
Finite-size scaling of the superfluid fraction for different values 
of 
the three-body cutoff $R_0 \rho^{1/3}$. 
The crossing point of 
$N^{1/3}\rho_s/\rho$ \emph{vs.} $T / T_c^0$
establishes a $10\%$ decrease of the superfluid transition temperature with 
respect to ideal bosons, in the limit $N \to \infty$ (\emph{cf.} Fig.\,4(c)).
\label{suppfig:finite_size_scaling} }
\end{figure}

The critical temperature $T_c$ is extracted from finite-$N$ data
using the scaling ansatz of Ref.\,\cite{Pollock1992PRB}. This assumes that in 
the critical region the
rescaled superfluid fraction $N^{1/3} \rho_s/\rho$ depends on the
system size $N$ only through the quantity $N^{1/(3\nu)} (T-T_c)/T_c$, where
$\nu$ is the correlation-length critical exponent, and 
implies that $N^{1/3} \rho_s/\rho$ becomes size-independent at the critical
temperature $T=T_c$ of the infinite system.
The dependence of $N^{1/3} \rho_s/\rho$ on system size,
for different values of the three-body cutoff $R_0 \rho^{1/3}$,
is shown in \fig{suppfig:finite_size_scaling},
and we observe that the crossing point is roughly at 90\% of
the critical temperature of ideal bosons\cite{Pitaevskii2003}.
The critical temperature $T_c$ weakly depends on $R_0 \rho^{1/3}$:
In the range $0.164 \lesssim R_0\rho^{1/3} \lesssim 0.204$, it
increases from $T_c/T_c^0\approx0.87$ to $T_c/T_c^0\approx 0.91$.

\subsection{Supplementary Item 4: Effect of interaction on $n(\kvec)$}
\begin{figure}[hbt]
\centering
\includegraphics[width=0.5\linewidth]{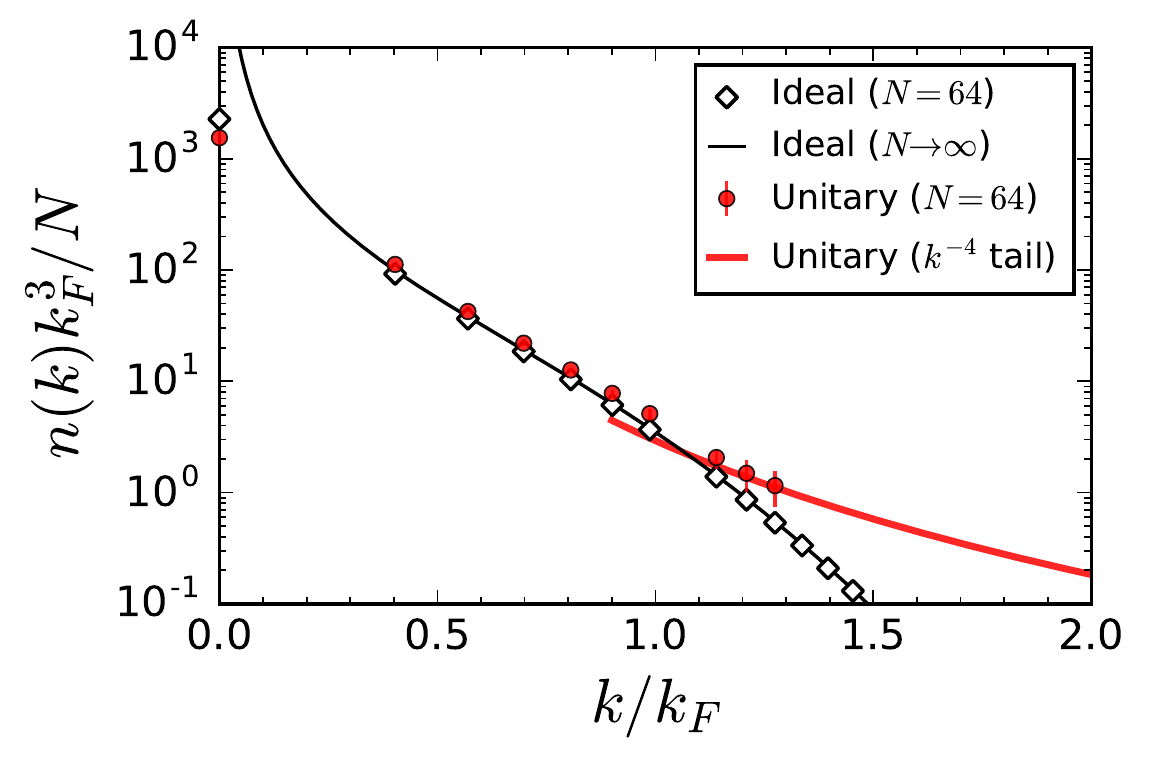}
\caption{
Momentum distribution $n(\kvec)$ in the Bose-condensed phase (point D in
Fig.\,3(a)) for the unitary Bose gas with $N=64$ (same data as in
Fig.\,4(a)), compared to the curves for finite and infinite systems of
ideal bosons at the same temperature.
\label{suppfig:nk_ideal} }
\end{figure}

In the Bose-condensed phase, the $\kvec=\mathbf{0}$ component of the momentum
distribution is reduced by unitary interactions, and the presence of a
slowly-decaying $k^{-4}$ tail at large $k$ does not fully compensate this
decrease.
Therefore, the unitary-gas momentum distribution has a stronger weight
in the intermediate-$k$ region, as clearly visible in \fig{suppfig:nk_ideal}.
In both the interacting and non-interacting case, $n(\kvec)$ does not show
strong finite-size effects at $k>0$.

\subsection{Supplementary Item 5: Coexistence free energy}
We consider $N$ particles in a fixed volume $V$, in the presence of an
$l$-particle nucleus of Efimov liquid.
The coexistence free energy includes the liquid and gas contributions.
For the liquid, we approximate $F_\mathrm{liq}(l)\simeq E_\mathrm{liq}(l)$,
neglecting the entropic contribution, and we use the cluster energies from
Ref.\,\cite{VonStecher2010JPB} for  $l \leq 13$, in terms of the trimer energy
$|E_T| \simeq 0.00214\,\hbar^2 / (m R_0^2)$ \cite{Braaten2006PR}.
For the gas contribution, we consider $N-l$ particles in a volume
$V-V_\mathrm{liq}$ (where $V_\mathrm{liq}\simeq l\times(3.65 R_0)^3$), and
compute
$F_\mathrm{gas}(N-l)$ up to the third virial coefficient\cite{Castin2013CJP}.
At given values of $N,V,$ and $T$, the coexistence free energy reads
\begin{equation}
F^N_\mathrm{coex}(l) \simeq E_\mathrm{liq}(l) + F_\mathrm{gas}(N-l).
\end{equation}
Computing $F^N_\mathrm{coex}(l)$ as a function of $l$ allows us to check for 
the existence of a free-energy barrier $\beta \Delta F$, which does not
disappear in the low-temperature regime.
The third-order virial and cluster expansions differ in their range of 
validity, the cluster expansion being more accurate at low temperature
(\emph{cf.} Fig.\,3(c)).
We find that the above model is not quantitatively reliable at large $R_0$, for
which the instability takes place at lower temperature. Its limit of validity is
$R_0\rho^{1/3}\lesssim 0.04$, while for larger values of $R_0$ it does not
correctly reproduce the coexistence line.

\end{document}